\documentclass[twocolumn,prl,preprintnumbers]{revtex4-1}
\usepackage{amsmath,amssymb,graphicx,booktabs,bm,psfrag,color,slashed,euscript,mathtools}
\newcommand{\spac}{{\hspace{0.3mm}}}
\newcommand{\braces}[1]{[\hspace{-0.5mm}[#1]\hspace{-0.5mm}]}

\begin{document}

\title{Factorization at Subleading Power, Sudakov Resummation and\\ 
Endpoint Divergences in Soft-Collinear Effective Theory}

\preprint{MITP/20-050, ZU-TH 31/20}
\preprint{September 9, 2020}
% arXiv:2009.04456
% v1: 9 September 2020 (revised: 20 November 2020, 20 April 2021)
% v2: 16 June 2021

\author{Ze Long Liu$^a$}
\author{Bianka Mecaj$^b$}
\author{Matthias Neubert$^{b,c,d}$}
\author{Xing Wang$^b$}

\affiliation{${}^a$Theoretical Division, Los Alamos National Laboratory, Los Alamos, NM 87545, U.S.A.\\ 
${}^b$PRISMA$^+$\! Cluster of Excellence {\rm \&} Mainz Institute for Theoretical Physics, Johannes Gutenberg University, 55099 Mainz, Germany\\
${}^c$Department of Physics {\em\&} LEPP, Cornell University, Ithaca, NY 14853, U.S.A.\\
${}^d$Department of Physics, Universit\"at Z\"urich, Winterthurerstrasse 190, CH-8057 Z\"urich, Switzerland}

\begin{abstract}
Starting from the first renormalized factorization theorem for a process described at subleading power in soft-collinear effective theory, we discuss the resummation of Sudakov logarithms for such processes 
in renormalization-group improved perturbation theory. Endpoint divergences in convolution integrals, which arise generically beyond leading power, are regularized and removed by systematically rearranging the factorization formula. We study in detail the example of the $b$-quark induced $h\to\gamma\gamma$ decay of the Higgs boson, for which we resum large logarithms of the ratio $M_h/m_b$ at next-to-leading logarithmic order. We also briefly discuss the related $gg\to h$ amplitude.
\end{abstract}

\maketitle

\subsection{I.\,~Introduction}
\vspace{-1mm}

Soft-collinear effective theory (SCET) \cite{Bauer:2001yt,Bauer:2002nz,Beneke:2002ph} provides an efficient framework for addressing the problem of scale separation for cross sections and decay rates in high-energy physics. In order to fully establish SCET as a systematic and versatile tool and apply it to several observables of phenomenological interest, it is important to understand its structure beyond the leading order in power counting. Indeed, much recent work has aimed at exploring factorization theorems at subleading power -- a problem that turns out to be unexpectedly intricate and subtle. Specific applications considered include the study of power corrections to event shapes \cite{Moult:2016fqy,Moult:2019vou} and transverse-momentum distributions \cite{Ebert:2018gsn}, the threshold factorization for the Drell-Yan process \cite{Bonocore:2015esa,Beneke:2018gvs,Bahjat-Abbas:2019fqa,Beneke:2019oqx}, and power-suppressed contributions to Higgs-boson decays \cite{Liu:2019oav,Wang:2019mym}. One finds that technical complications arise which do not occur at leading power. The most puzzling one is the appearance of endpoint-divergent convolution integrals over products of component functions, each depending on a single scale \cite{Moult:2019mog,Beneke:2019kgv,Moult:2019uhz,Beneke:2019oqx,Moult:2019vou,Liu:2019oav,Wang:2019mym,Beneke:2020ibj,Liu:2020wbn}. While several of these studies derived factorized expressions for cross sections or decay rates in terms of convolutions of bare matching coefficients with SCET matrix elements, in most cases the presence of endpoint divergences prevented the establishment of a proper factorization formula in terms of renormalized objects. 

\begin{figure}[t]
\begin{center}
\vspace{1mm}
\includegraphics[width=0.35\textwidth]{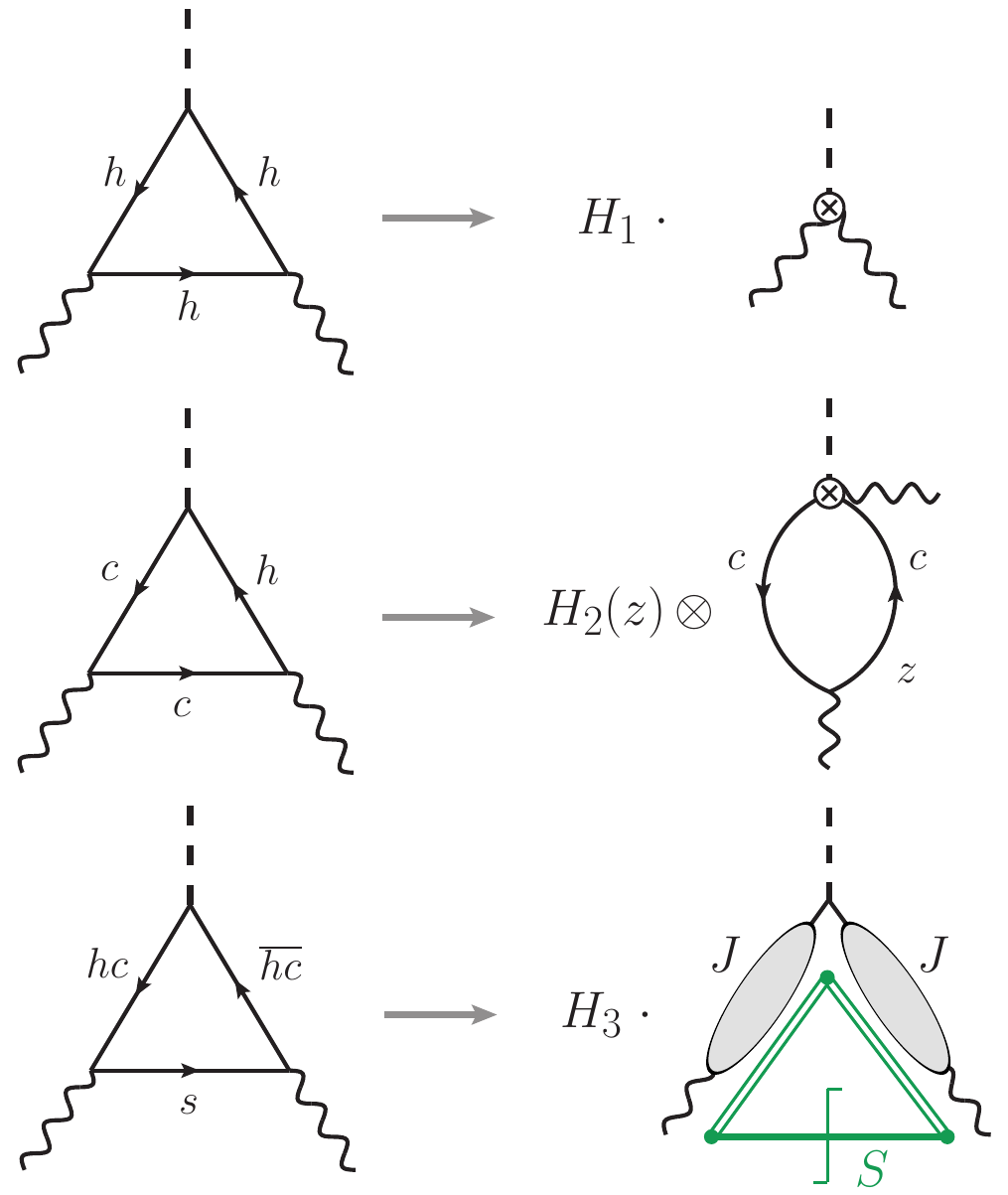} 
\caption{\label{fig:scet1} 
Leading regions of loop momenta contributing to the decay amplitude. The convolution symbol $\otimes$ in the second term means an integral over $z$.}
\vspace{-6mm}
\end{center}
\end{figure}

In \cite{Liu:2019oav,Liu:2020wbn} we have initiated a detailed discussion of SCET factorization at subleading power for processes subject to rapidity divergences. As a concrete example we have factorized the decay amplitude for the radiative Higgs-boson decay $h\to\gamma\gamma$ mediated by the Higgs coupling to bottom quarks. This pseudo-observable starts at subleading power in the SCET expansion. The absence of a leading-order contributions ensures that relatively few SCET operators appear in the factorization theorem, and we have succeeded to derive the non-local renormalization-group evolution equations (RGEs) for the various SCET operators and Wilson coefficients at one and (in part) two-loop order. The $h\to\gamma\gamma$ decay amplitude receives large logarithms of the form $\alpha\spac\alpha_s^n L^k$, where 
% $\alpha$ and $\alpha_s$ are the QED and QCD couplings, respectively, and  
$L=\ln(-M_h^2/m_b^2-i0)$ and $k\le 2n+2$. In order to resum these logarithms it is necessary to factorize the amplitude into objects depending only on one of the three relevant scales set by the Higgs-boson mass $M_h$, the mass $m_b$ of the bottom quark and the intermediate scale $\sqrt{M_h\spac m_b}$. From a phenomenological point of view, the contributions to the $h\to\gamma\gamma$ amplitude mediated by bottom and charm quarks, to which our approach can be applied, account for about 1.6\% of the decay rate (at leading order and using quark pole masses), while the analogous contributions to $gg\to h$ production lower the rate by approximately 13\%. In extensions of the Standard Model with enhanced Yukawa couplings of the bottom and charm quarks, these contributions can be much enhanced (see e.g.\ \cite{Spira:1995rr}). The development of a formalism that allows for a consistent resummation of Sudakov logarithms for subleading-power observables is therefore important for high-precision predictions of Higgs-boson production and decay in the Standard Model and beyond.

The ``bare'' factorization formula for the decay amplitude contains three terms consisting of unrenormalized SCET operators multiplied by bare Wilson coefficients, which account for the hard matching corrections arising when the ``full theory'' (the Standard Model with top quarks integrated out) is matched onto SCET. In its simplest form, the factorization formula reads 
\begin{align}\label{barefact}
   & {\cal M}_b = H_1^{(0)} \langle O_1^{(0)}\rangle 
    + 2\int_0^1\!\!dz\,H_2^{(0)}\!(z)\,\langle O_2^{(0)}\!(z)\rangle \\
   &\! + H_3^{(0)}\!\!\!\int\!\!\!\!\int_0^\infty\!\!\frac{d\ell_+}{\ell_+} \frac{d\ell_-}{\ell_-}\,
    J^{(0)}\!(-M_h\ell_+)\,J^{(0)}\!(M_h\ell_-)\,S^{(0)}\!(\ell_+\ell_-) \spac , \notag
\end{align}
where the $h\to\gamma\gamma$ matrix elements are evaluated on shell. The three terms correspond to different regions of loop momenta giving rise to leading contributions to the decay amplitude ${\cal M}_b$ (with the photon polarization vectors removed), as illustrated in Figure~\ref{fig:scet1}. The operator $O_1$ contains a Higgs field coupled to two collinear gauge fields describing photons moving along opposite light-like directions $n$ and $\bar n$. It descents from full-theory graphs in which all internal momenta are hard, of order $M_h$. The operator $O_2(z)$ contains a Higgs field, an $\bar n$-collinear photon field and two $n$-collinear $b$-quark fields, which share the momentum of the $n$-collinear photon. The variable $z$ denotes the longitudinal momentum fraction carried by one of the quarks. This operator is generated by full-theory graphs in which a loop momentum is collinear with the photon direction $n$ and carries virtuality of order $m_b$. The factor~2 in front of this term arises because there is an analogous contribution with $n$ and $\bar n$ interchanged. Finally, the third operator consists of the time-ordered product of the scalar Higgs current with two subleading-power terms in the SCET Lagrangian, in which hard-collinear fields are coupled to a soft quark field. It arises from full-theory graphs containing a soft quark propagator between the two photons, with all momentum components of order $m_b$. The other quark propagators are then off-shell with virtualities of order $\sqrt{M_h m_b}$. Because of this scale hierarchy, the $h\to\gamma\gamma$ matrix element of this operator can be factorized further into a convolution of two jet functions with a soft function, as shown in (\ref{barefact}). 

Major complications arise from endpoint-divergent convolution integrals in the second and third term in (\ref{barefact}), which need to be properly identified and regularized. The presence of a soft-quark contribution has been identified as the source of these endpoint divergences. The integral over $z$ in the second term contains singularities at $z=0$ and $z=1$, because at lowest order in perturbation theory the Wilson coefficient $H_2^{(0)}\propto[z(1-z)]^{-1}$ while the matrix element $\langle O_2^{(0)}\rangle$ is $z$ independent. Likewise, the integrals over $\ell_+$ and $\ell_-$ in the third term contain singularities for $\ell_\pm\to\infty$, since at lowest order the jet and the soft functions are given by constants. In higher orders, some of these endpoint divergences are regularized by the dimensional regulator $D=4-2\epsilon$, but others require an additional rapidity regulator \cite{Becher:2010tm,Chiu:2012ir,Li:2016axz}. In \cite{Liu:2019oav} we have regularized the rapidity divergences by means of an analytic regulator imposed on the convolution variables $z$ and $\ell_\pm$. The singular contributions in the rapidity regulator cancel in the sum of the second and third term of the factorization formula. This requires that for $z\to 0$ or 1 these two terms must have closely related structures, which is ensured by two exact, $D$-dimensional refactorization conditions, which have been proven using SCET methods in \cite{Liu:2020wbn}. With the help of these relations we have recast the factorization formula in such a way that the singularities in the second term are removed by subtractions of the integrand. 

In this Letter, we use the formalism developed in \cite{Liu:2020wbn} to perform the resummation of large Sudakov logarithms for the $h\to\gamma\gamma$ decay amplitude beyond the leading-logarithmic approximation. This is the first time such a resummation has been accomplished in SCET. We begin by briefly recalling the derivation of the renormalized factorization formula. We then focus on the term in the formula that contains the leading logarithmic (LL) and next-to-leading logarithmic (NLL) contributions and perform analytically the resummation of logarithmic corrections at leading order in RG-improved perturbation theory. From this result, we derive the infinite tower of LL and NLL logarithms. We then briefly discuss the extension of our approach to the related $gg\to h$ process, showing that the presence of colored particles in the initial state does not pose any new difficulties.

\subsection{II.\,~Factorization with endpoint divergences}
\vspace{-3mm}

The main accomplishment of our work \cite{Liu:2020wbn} was the derivation of a {\em renormalized\/} factorization formula for the $b$-quark induced $h\to\gamma\gamma$ amplitude, in which all bare quantities are replaced by their renormalized counterparts. The result reads (with $\bar z\equiv 1-z$)
\begin{widetext}
\begin{equation}\label{factren}
\begin{aligned}
   {\cal M}_b
   &= H_1(\mu)\,\langle O_1(\mu)\rangle 
    + 2 \int_0^1\!dz\,\bigg[ H_2(z,\mu)\,\langle O_2(z,\mu)\rangle 
    - \braces{H_2(z,\mu)}\,\braces{\langle O_2(z,\mu)\rangle} 
    - \braces{H_2(\bar z,\mu)}\,\braces{\langle O_2(\bar z,\mu)\rangle} \bigg] \\
   &\quad + \lim_{\sigma\to-1}\,H_3(\mu) 
    \int_0^{M_h}\!\frac{d\ell_-}{\ell_-}\,\int_0^{\sigma M_h}\!\frac{d\ell_+}{\ell_+}\,    
    J(M_h\ell_-,\mu)\,J(-M_h\ell_+,\mu)\,S(\ell_+\ell_-,\mu) \,\Big|_{\rm leading\,\,power} \,,
\end{aligned}
\end{equation}
\end{widetext}
which is free of endpoint divergences. The symbol $\braces{f(z)}$ means that one retains only the leading terms of a function $f(z)$ in the limit $z\to 0$ and neglects higher power corrections. The limit $\sigma\to-1$ in the last term must be taken by analytic continuation. Note the important fact that the hard cutoffs in the third term are not put in by hand, but are an unavoidable consequence of eliminating the endpoint divergences in the second term by means of plus-type subtractions. The presence of these cutoffs breaks the homogeneous power counting of the SCET matrix elements, and only the leading-power contributions to the last term should be kept for consistency. This effect is a manifestation of the collinear anomaly, the fact that a classical symmetry of SCET under rescalings of the light-cone vectors $n$ and $\bar n$ is broken by quantum effects \cite{Becher:2010tm}. As we will see later, the presence of the cutoffs gives rise to a highly non-trivial structure of large logarithmic corrections to the decay amplitude. 

Establishing relation (\ref{factren}) has been non-trivial, because the presence of cutoffs on some of the convolution integrals does not commute with renormalization. For example, the renormalization condition for the soft function reads $S(w,\mu)=-\int_0^\infty\!dw'\,Z_S(w,w')\,S^{(0)}(w')$ \cite{Liu:2020eqe}, and an analogous equations holds for the jet function $J$ \cite{Bosch:2003fc,Liu:2020ydl}. Moving the cutoffs from the bare to the renormalized functions gives rise to extra terms, which individually have a rather non-trivial structure. We have shown that, to all orders of perturbation theory, the {\em sum\/} of the extra terms depends on the high scale $M_h$ only and can be absorbed into the renormalization condition for the hard matching coefficient $H_1^{(0)}$.

While the derivations in \cite{Liu:2019oav,Liu:2020wbn} have focused on one particular observable, the method developed there, i.e.\ the removal of endpoint divergences using plus-type subtractions and $D$-dimensional refactorization conditions, is more general and can be applied to other observables as well. As an important example, we will discuss the case of Higgs production in gluon-gluon fusion, which is more complicated due to the presence of colored particles in the initial state.

\subsection{III.\,~Sudakov resummation at subleading power and NLL order}
\vspace{-1mm}

In \cite{Liu:2020eqe,Liu:2020ydl,Liu:2020wbn} we have derived the explicit form of the RGEs obeyed by all quantities entering the factorization formula (\ref{factren}). It follows from the structure of these equations that, if the factorization scale $\mu$ is chosen of order the hard scale $M_h$, both the LL and NLL corrections to the decay amplitude are contained in the last term, $T_3$, shown in the second line of (\ref{factren}). This contribution is enhanced, because the integrals over $\ell_\pm$ produce two powers of large rapidity logarithms. As mentioned earlier, this is a consequence of the collinear anomaly \cite{Becher:2010tm}. In previously studied examples where the collinear anomaly appears the rapidity logarithms take on a simpler form and (typically) exponentiate. In the present case their structure is more complicated, because they arise from a double integral over a rather complicated integrand. Previous authors have resummed the series of the leading double logarithms of order $\alpha\spac\alpha_s^n L^{2n+2}$ \cite{Kotsky:1997rq,Akhoury:2001mz,Liu:2017vkm,Liu:2018czl}, which are contained in $T_3$. These resummations were not based on a factorization formula, but on a relation of the leading logarithmic contributions to the $h\to\gamma\gamma$ amplitude (those arising from soft-quark exchange) with the off-shell Sudakov form factor studied in \cite{Smilga:1979uj}. For practical applications, it is however important to go beyond the LL approximation and perform the resummation in RG-improved perturbation theory. Only then it is ensured that all large logarithms are exponentiated. Here we illustrate this for the case of $T_3$. 

At next-to-leading order in QCD perturbation theory, the explicit expressions for the hard, jet and soft functions contain logarithmic contributions involving different scales. The hard function $H_3$ is given by
\begin{equation}
   H_3(\mu) = \frac{y_b(\mu)}{\sqrt 2} \left[ - 1 + \frac{\alpha_s}{3\pi}
    \left( \ln^2\frac{-M_h^2-i0}{\mu^2} + 2 - \frac{\pi^2}{6} \!\right) \right] ,
\end{equation}
where $y_b(\mu)$ denotes the running $b$-quark Yukawa coupling. For the jet function one obtains \cite{Bosch:2003fc}
\begin{equation}\label{Jres}
   J(p^2,\mu) = 1 + \frac{\alpha_s}{3\pi} \left( \ln^2\frac{-p^2-i0}{\mu^2}%\bigg)
    - 1 - \frac{\pi^2}{6} \right) .
\end{equation}
Finally, the renormalized soft function is given by \cite{Liu:2020eqe}
\begin{equation}\label{Swres}
   S(w,\mu) = - \frac{\alpha}{3\pi}\,m_b(\mu) \left\{
    \begin{array}{l}
     S_a(w,\mu) \,; \quad w > m_b^2 \,, \\[1mm]
     S_b(w,\mu) \,; \quad w < m_b^2 \,,
   \end{array} \right.
\end{equation}
where in the prefactor $m_b(\mu)$ is the running $b$-quark mass, while the two cases are differentiated by whether the variable $w$ is larger or smaller than the $b$-quark pole mass $m_b$. Specifically, one finds (for $\hat w\equiv w/m_b^2$)
\vspace{-1mm}
\begin{align}\label{Sres}
   S_a(w,\mu) &= 1 + \frac{\alpha_s}{3\pi} \Big[ - L_w^2 - 6 L_w + 12 - \frac{\pi^2}{2} 
    + g(\hat w) \Big] \,, \nonumber\\
   S_b(w,\mu) &= \frac{4\alpha_s}{3\pi}\,\ln(1-\hat w)\,\big[ L_m + \ln(1-\hat w) \big] \,.    
\end{align}
where $L_w=\ln(w/\mu^2)$, $L_m=\ln(m_b^2/\mu^2)$, and the function $g(\hat w)$ vanishes for $\hat w\to\infty$. The soft function vanishes linearly for $w\to 0$. It is evident from the above expression that there is no single choice of the factorizations scale $\mu$ for which all three functions are free of large logarithms.

The RGEs for the jet and soft functions have been derived up to two-loop order in \cite{Bosch:2003fc,Liu:2020ydl,Liu:2020eqe}. Choosing a high value $\mu=\mu_h\sim M_h$ for the factorization scale (see above), and using the explicit solutions of the RGEs for $J$ and $S$ derived in these works, we have obtained a closed analytic expression for $T_3$ at leading order (LO) in RG-improved perturbation theory. It reads (omitting for simplicity the $-i0$ prescription in the last three factors of the first line) 
\begin{widetext}
\begin{equation}\label{T3resummed}
\begin{aligned}
   T_3^{\rm LO} &= \frac{\alpha}{3\pi}\spac\frac{y_b(\mu_h)}{\sqrt 2}\! 
    \int_0^{M_h}\!\frac{d\ell_-}{\ell_-} \int_0^{M_h}\!\frac{d\ell_+}{\ell_+}\,m_b(\mu_s)\,
    e^{2{\cal S}_F(\mu_s,\mu_h)-2{\cal S}_F(\mu_-,\mu_h)-2{\cal S}_F(\mu_+,\mu_h)}\! 
    \left( \frac{-M_h\ell_-}{\mu_-^2} \right)^{\!a_\Gamma^-}\!\!\!
    \left( \frac{-M_h\ell_+}{\mu_+^2} \right)^{\!a_\Gamma^+}\!\!\!
    \left( \frac{-\ell_+\ell_-}{\mu_s^2} \right)^{\!-a_\Gamma^s} \\
   &\quad\times \left( \frac{\alpha_s(\mu_s)}{\alpha_s(\mu_h)} \right)^{\!-\frac{\gamma_{s,0}}{2\beta_0}} 
    e^{-2\gamma_E\spac a_\Gamma^+}\,\frac{\Gamma(1-a_\Gamma^+)}{\Gamma(1+a_\Gamma^+)}\,
    e^{-2\gamma_E\spac a_\Gamma^-}\,\frac{\Gamma(1-a_\Gamma^-)}{\Gamma(1+a_\Gamma^-)}\,
    e^{4\gamma_E\spac a_\Gamma^s}\,
    G_{4,4}^{2,2} \left( 
     \begin{array}{cccc} -a_\Gamma^s\,,\!\! & -a_\Gamma^s\,,\!\! & 1\!-\!a_\Gamma^s\,,\!\! 
     & 1\!-\!a_\Gamma^s \\
     0\,,\!\! & 1\,,\!\! & 0\,,\!\! & 0 \end{array} \bigg| \frac{m_b^2}{-\ell_+\ell_-} \right) .
\end{aligned}
\end{equation}
\end{widetext}
The Sudakov exponent ${\cal S}_F$ and the exponents $a_\Gamma^i$ are given by \cite{Becher:2006mr}
\begin{align}
   {\cal S}_F(\mu_i,\mu_h) 
   &= \frac{C_F\spac\gamma_0^{\rm cusp}}{4\beta_0^2}\,\bigg[
    \frac{4\pi}{\alpha_s(\mu_i)} \left( 1 - \frac{1}{r} - \ln r \right) \notag\\
   & + \left( \frac{\gamma^{\rm cusp}_1}{\gamma^{\rm cusp}_0} - \frac{\beta_1}{\beta_0} \right) 
    \left( 1 - r + \ln r \right) + \frac{\beta_1}{2\beta_0} \ln^2 r \Bigg] \spac , \notag\\
    a_\Gamma(\mu_i,\mu_h)
    &= \frac{C_F\spac\gamma^{\rm cusp}_0}{2\beta_0}\,\ln\frac{\alpha_s(\mu_h)}{\alpha_s(\mu_i)} 
     \equiv a_\Gamma^i \,,
\end{align}
where $r=\alpha_s(\mu_h)/\alpha_s(\mu_i)$, $\gamma^{\rm cusp}_{0,1}$ are the one- and two-loop coefficients of the cusp anomalous dimension \cite{Becher:2009qa}, $\beta_{0,1}$ are the corresponding coefficients of the QCD $\beta$-function, and $\gamma_{s,0}=-6\spac C_F$ is the one-loop anomalous dimension of the soft function \cite{Liu:2020eqe}. The object $G_{4,4}^{2,2}(\cdots|x)$ is a Meijer $G$-function (see e.g.\ \cite{Gfun2,Gfun3}). This function vanishes for $|x|\to\infty$ (and hence the region where $\ell_+\ell_-\ll m_b^2$ gives a power-suppressed contribution to $T_3$), whereas it approaches $\Gamma^2(1+a_\Gamma^s)/\Gamma^2(1-a_\Gamma^s)$ for $x\to 0$. 

The matching scales $\mu_\pm$ for the jet functions and $\mu_s$ for the soft function must be chosen such that the matching conditions at these scales are free of large logarithms. This is a non-trivial requirement, because the jet and soft functions depend on the variables $\ell_\pm$, which are integrated from the soft region ($\ell_+\ell_-\sim m_b^2$) into the hard region ($\ell_+\ell_-\sim M_h^2$). It is therefore necessary that one sets the matching scales {\em dynamically\/} under the integral \cite{Liu:2020eqe}, such that $\mu_s^2\sim \ell_+\ell_-$ and $\mu_\pm^2\sim M_h\ell_\pm$ up to ${\cal O}(1)$ factors, see (\ref{Jres}) and (\ref{Sres}). (The three scale parameters should however not be lowered below $m_b^2$, because the region where $\ell_+\ell_-$ is parametrically smaller than $m_b^2$ gives a power-suppressed contribution to the decay amplitude.) When this is done, all large logarithms are resummed into ratios of running couplings. Corrections omitted in (\ref{T3resummed}) are thus suppressed by powers of $\alpha_s$. They can be included systematically by calculating the matching conditions and anomalous dimensions to higher orders. For the method of dynamical scale setting to be consistent, it is important that the renormalized soft function $S(w,\mu)$ in (\ref{Swres}) does not develop large logarithms of the form $\ln^n(w/m_b^2)$ in the limit where $w\gg m_b^2$ \cite{Liu:2020wbn}, because otherwise it would need to be refactorized when the variables $\ell_\pm$ in (\ref{factren}) are in the hard region. This property of the soft function follows from the fact that its RGE is independent of the $b$-quark mass \cite{Liu:2020eqe}.

Equation (\ref{T3resummed}) is the main result of this work, which accomplishes for the first time the resummation of Sudakov logarithms for a subleading-power quantity in RG-improved perturbation theory. As a cross check, we have reexpanded the resummed expression (\ref{T3resummed}) in a perturbative series and extracted the infinite tower of LL and NLL contributions of order $\alpha\spac\alpha_s^n L^k$ with $k=2n+2$ and $k=2n+1$. We find (with $C_F=\frac43$ and $\beta_0=11-\frac23\spac n_f$)
\begin{equation}\label{NLLres}
\begin{aligned}
   {\cal M}_b^{\rm NLL} 
   &= \frac{\alpha}{3\pi}\,m_b\,\frac{y_b(\hat\mu_h)}{\sqrt 2}\,\frac{L^2}{2}
    \sum_{n=0}^\infty \left( - \rho \right)^n \frac{2\spac\Gamma(n+1)}{\Gamma(2n+3)} \\
   &\quad\!\times \left[ 1 + \frac{3\rho}{2L}\,\frac{2n+1}{2n+3}
    - \frac{\beta_0}{C_F}\,\frac{\rho^2}{4L}\,\frac{(n+1)^2}{(2n+3)(2n+5)} \right]\! ,
\end{aligned}
\end{equation}
where $\rho=\frac{C_F\spac\alpha_s(\hat\mu_h)}{2\pi}\spac L^2$ with $L=\ln(-M_h^2/m_b^2-i0)$ and $\hat\mu_h^2=-M_h^2-i0$, and $m_b$ denotes the pole mass. Our normalization of the amplitude is chosen such that this result can be compared directly with the findings of \cite{Akhoury:2001mz}. We observe a disagreement in the second term in brackets, which is quoted in this reference as $\frac{3\rho}{2L}\,\frac{n+1}{2n+3}$. The infinite sums in (\ref{NLLres}) can be expressed in closed form in terms of a hypergeometric function and the Dawson integral.

\subsection{\boldmath IV.\,~Factorization and resummation of the\\ $gg\to h$ amplitude}
\vspace{-3mm}

It is straightforward to apply the formalism developed in \cite{Liu:2020wbn} to the closely related $b$-quark induced contributions to the $gg\to h$ production process \cite{inprep}. The different contributions to the decay amplitude and the corresponding SCET operators have the same form as shown in Figure~\ref{fig:scet1} (with photon fields replaced by gluon fields), and the renormalized factorization theorem is of the same form as in (\ref{factren}). The presence of colored particles in the external states does not invalidate our approach of dealing with the endpoint divergences. However, it implies that the $gg\to h$ amplitude by itself is not an infrared-safe quantity. Soft and collinear emissions from the initial-state gluons give rise to additional $1/\epsilon^n$ poles, which must be factored off and absorbed into the renormalization of the gluon distribution functions. This is accomplished by means of a global renormalization factor in the $\overline{\rm MS}$ scheme \cite{Becher:2009qa,Becher:2009cu}, such that
\begin{equation}
   {\cal M}_{gg}(\mu) = Z_{gg}^{-1}(\mu)\,{\cal M}_{gg}^{(0)} \,,
\end{equation} 
where ${\cal M}_{gg}^{(0)}$ refers to the bare $gg\to h$ production amplitude. The renormalized amplitude carries an overall scale dependence, which compensates the $\mu$ dependence of the parton distribution functions and the relevant soft function. At LO in RG-improved perturbation theory, we find 
\begin{equation}
  {\cal M}_{gg}(\mu) 
  = e^{2{\cal S}_A(\hat\mu_h,\mu)}\,\frac{\alpha_s(\mu)}{\alpha_s(\hat\mu_h)}\,{\cal M}_{gg}(\hat\mu_h) \,,
\end{equation}
where the quantity ${\cal S}_A$ is obtained by replacing $C_F\to C_A$ in the expression for ${\cal S}_F$ given above. The Sudakov resummation for the quantity ${\cal M}_{gg}(\hat\mu_h)$ can be performed in a similar way as for the $h\to\gamma\gamma$ case, by generalizing the RGEs for the soft and jet functions to the non-abelian case \cite{inprep}. In analogy with the result (\ref{NLLres}), we obtain
\begin{equation}\label{ggHNLLres}
\begin{aligned}
  &{\cal M}_{b,gg}^{\rm NLL}(\hat\mu_h) 
   = \delta_{AB}\,\frac{\alpha_s(\hat\mu_h)}{2\pi}\,m_b\,\frac{y_b(\hat\mu_h)}{\sqrt 2}\,
   \frac{L^2}{2} \\
  &\quad\times \sum_{n=0}^\infty \left(-\rho_g\right)^n \frac{2\spac\Gamma(n+1)}{\Gamma(2n+3)}\,
   \bigg[ 1 + \frac{C_F}{C_F-C_A}\,\frac{3\rho_g}{2L}\,\frac{2n+1}{2n+3} \\
  &\hspace{1.2cm} - \frac{\beta_0}{C_F-C_A}\,\frac{\rho_g^2}{4L}\,\frac{(n+1)^2}{(2n+3)(2n+5)} \bigg] \,,
\end{aligned}
\end{equation}
where $\rho_g=\frac{(C_F-C_A)\spac\alpha_s(\hat\mu_h)}{2\pi}\spac L^2$, and $A,B$ are the color indices of the gluons. Remarkably, the series of LL and NLL terms in the ``abelian process'' $h\to\gamma\gamma$ and the ``non-abelian process'' $gg\to h$ are related to each other by a simple replacement of color factors. For the LL contributions this was first shown in \cite{Liu:2017vkm}, and the result is extended here to NLL order. The series of NLL corrections has recently also been studied using non-SCET methods \cite{Anastasiou:2020vkr}. In the Erratum to their paper, the authors present a formula which agrees with (\ref{ggHNLLres}).

\subsection{V.\,~Conclusions and outlook}
\vspace{-3mm}

Based on a renormalized SCET factorization theorem for an observable appearing at subleading power in the ratio of two hierarchical mass scales $m\ll M$, in which endpoint-divergent convolution integrals are regularized by plus-type subtractions and the use of exact $D$-dimensional refactorization conditions, we have performed the first resummation of Sudakov logarithms for a subleading-power quantity ($T_3$) in RG-improved perturbation theory. We have focussed on the example of the $b$-quark induced $h\to\gamma\gamma$ decay of the Higgs boson, and we have briefly discussed the ``non-abelian'' extension to the case of Higgs production in gluon-gluon fusion. Re-expanding our main result (\ref{T3resummed}) in a perturbative series, we have extracted the infinite towers of leading and next-to-leading logarithmic corrections to the $h\to\gamma\gamma$ and $gg\to h$ amplitudes, finding that results for the subleading terms obtained using other methods need to be corrected. The techniques we have developed -- the elimination of endpoint divergences using a rearrangement of the bare factorization formula based on refactorization conditions, and the resummation of large logarithms using dynamical scale setting -- are more general and can be applied to many other power-suppressed observables as well. Important examples include the non-abelian extension of our work to Higgs production in gluon fusion \cite{inprep}, as well as power corrections to transverse-momentum spectra in Drell-Yan and Higgs production, jet-veto cross sections, the pion form factor, exclusive nonleptonic two-body decays of $B$ mesons, and others. Our methods are likely also needed to understand subleading-power factorization for observables not susceptible to rapidity divergences. For example, recently refactorization conditions analogous to those derived in \cite{Liu:2019oav,Liu:2020wbn} were used to resum the leading double-logarithmic corrections present in the off-diagonal DGLAP splitting kernels for large $x$ \cite{Beneke:2020ibj}. Our results thus constitute an important step toward establishing a robust framework for studying SCET factorization and scale separation at subleading order in scale ratios.

\vspace{1mm}
{\em Acknowledgements:\/} 
One of us (M.N.) thanks Gino Isidori, the particle theory group at Zurich University and the Pauli Center for hospitality during a sabbatical stay. This research has been supported by the Cluster of Excellence PRISMA$^+$\! funded by the German Research Foundation (DFG) within the German Excellence Strategy (Project ID 39083149). The research of Z.L.L.\ is supported by the U.S.\ Department of Energy under Contract No.~DE-AC52-06NA25396, the LANL/LDRD program and within the framework of the TMD Topical Collaboration.

\end{document}